\documentclass[submission,copyright]{eptcs}
\providecommand{\event}{TTC 2011}
\usepackage{graphicx}

\title{Solving the TTC 2011 Reengineering Case with MOLA and Higher-Order
Transformations}

\author{Agris Sostaks
\and 
Elina Kalnina 
\and 
Audris Kalnins
\and 
Edgars Celms 
\and 
Janis Iraids
\institute{Institute of Mathematics and Computer Science, University of Latvia,
\\ Raina bulvaris 29, LV-1459, Riga, Latvia} 
\email{\{agris.sostaks, elina.kalnina, audris.kalnins, edgars.celms,
janis.iraids\} @lumii.lv} }

\def\titlerunning{Solving the TTC 2011 Reengineering Case with MOLA}
\def\authorrunning{A. Sostaks, E. Kalnina, A. Kalnins, E. Celms, J. Iraids}

\begin{document} 

\maketitle

\begin{abstract}
The Reengineering Case of the Transformation Tool Contest 2011 deals with
automatic extraction of state machine from Java source code. The transformation task involves \emph{complex, non-local matching} 
 of model elements. This paper contains the solution of the task using model transformation language MOLA.

The MOLA solution uses higher-order transformations (HOT-s) to generate a part
of the required MOLA program. The described HOT approach allows creating reusable, complex model transformation libraries for generic 
tasks without modifying an implementation of a model transformation language. Thus model transformation 
users who are not the developers of the language can achieve the desired
functionality more easily.
\end{abstract}

\section{Introduction}
A solution for the Reengineering Challenge case study \cite{programunderstandingcase} of the
Transformation Tool Contest
2011\footnote{\url{http://planet-research20.org/ttc2011/}} (TTC) has
been presented in this paper. Model transformation language MOLA
\cite{Kalnins2005} has been used to solve the task. The task is to create a
simple state machine model for a Java syntax graph model encoding a state machine with a set of coding conventions. The task consists of a core task and two extensions.

The core task is to build states and transitions. States are created from non-abstract Java classes that extend 
the class named \verb|State|. The transitions are encoded by method calls to the specific method 
named \verb|Instance()| of the next state returning the singleton instance of that state on 
which the \verb|activate()| method is called. Transition's trigger and
action attributes are extracted in the extensions of the task. Values for the attributes are dependent mainly on 
the type of the container within which a transition activation call occurs. Thus, the solution should 
deal with the main challenge of the task - \emph{complex, non-local matching} of
model elements.

The task  has been largely designed so, that for Java elements meeting certain
criteria a parent or a child of the specific type must be found. Since the
provided Java metamodel has a deep containment and inheritance hierarchy, 
there are lots of different navigation paths how the searched element may be reached. 
A higher-order transformation has been used to
generate MOLA procedures dealing with complex, non-local matching of model elements.

The paper is structured as follows. Section \ref{sec:mola} provides a short
overview of the MOLA language and tool. Section \ref{sec:hot} describes the
higher-order transformation used by the solution. Section \ref{sec:sol} contains a
description of the solution. The paper ends with discussion in Section \ref{sec:dis}. Detailed description  of the transformation definition (including \emph{source
code} - MOLA diagrams) can be found in Appendix.

\section{MOLA Language}
\label{sec:mola} 

MOLA is a graphical model transformation language which combines the declarative means for pattern specification 
and imperative control structures determining the order of transformation execution. The formal description of 
MOLA and also MOLA tool can be downloaded here - \url{http://mola.mii.lu.lv}. 

\begin{figure}[h]
\centering
\includegraphics[width=300px]{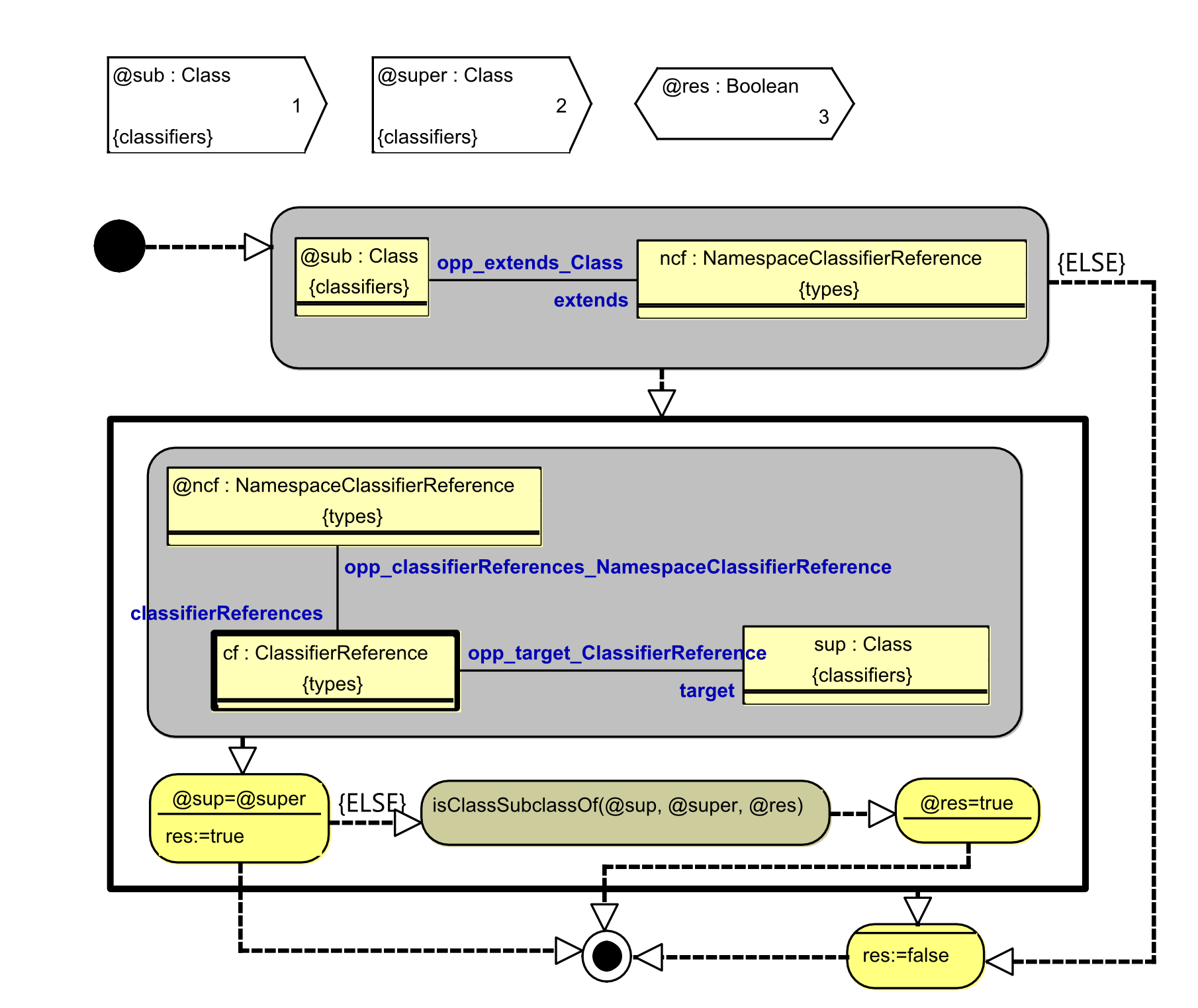}
\caption{The isClassSubclassOf procedure }
\label{fig:isclasssubclassof}
\end{figure}

The main element of MOLA transformation is a \emph{rule} (see the gray rounded rectangle in Figure \ref{fig:isclasssubclassof}). 
Rule contains a declarative pattern that specifies instances of which classes must be selected and how they must be linked. 
The instances to be included in the search are specified using \emph{class elements}.
Class elements may contain constraints defined using simple OCL-like expressions. 
Additionally, a rule may contain association links between class elements. Association links specify links of the exact type 
required to exist between the corresponding instances in a model.

In order to iterate through a set of instances MOLA provides a \emph{foreach loop} statement 
(See the bold rectangle in Figure \ref{fig:isclasssubclassof}). The
\emph{loophead} is a special kind of a rule that is used to specify the set of instances to be iterated over. The pattern is specified for the loophead 
in the same way as for ordinary rule. However, the \emph{loop variable} is a special class element 
(see the \verb|cf : ClassifierReference| element in Figure \ref{fig:isclasssubclassof}). 
A foreach loop is executed for each distinct instance that corresponds to the
loop variable and satisfies all constraints of the pattern.         

MOLA transformations can be compiled to several technical spaces (model repositories) - Eclipse Modeling Framework (EMF), 
JGraLab \cite{jgralab} and MII\_REP developed by IMCS, University of Latvia. Example models and metamodels in the 
reengineering challenge conform to the EMF technical space. 

Development of model transformations begins with importing source and target Ecore metamodels into MOLA Tool. 
The current version of MOLA requires all metamodel associations to be navigable
in both directions (this permits to perform an efficient pattern matching using
simple matching algorithms). Since a typical Ecore metamodel has many
associations navigable in one direction, the import facility has to extend the
metamodel - \emph{missing} opposite references are added automatically. The traceability links between the source and target model elements can be
added manually. In the given case, one association between the \verb|Class| and
\verb|State| classes is sufficient. MOLA execution environment (MOLA runner)
includes a generic model import facility, which automatically adjusts the imported model to the modified metamodel. 
Similarly, a generic export facility automatically strips all elements of the transformed 
model which does not correspond to the original target metamodel. 

Since the MOLA language does not have recursive patterns (as most graphical
languages), tasks like to answer whether a Java class is a subclass of another
class should be solved using a recursive procedure. A much more complex example for non-local matching is finding a Java class
method which contains the given expression. This task arises because the value
of a trigger attribute of a transition depends primarily on the class method and
secondarily on the statement (e.g. is it a switch case or catch block) within which the activation call occurs. The solution of the task also requires
a recursive approach, because the containment hierarchy of Java expressions and statements is deep and recursive. 
It means that every containment case should be specified in MOLA procedure as
 a separate rule thus producing large number of rules to be created.        

In total 22 distinct types and 44 distinct composition links should be checked
which results in 22 text statements and 44 rules in the MOLA procedure. 
However no element has been specified \emph{by hand} - the MOLA procedure has
been entirely generated. Of course, if some semantic constraints would 
be taken into account (e.g. the \verb|activate()| method is a void method and cannot be part of an \verb|and| or \verb|or| expression), 
the number of rules needed for MOLA would be much less. 
However, since these constraints haven't been stated in the task description, we have made a \emph{full} solution.

\section{Higher-Order Transformations using MOLA}
\label{sec:hot}

MOLA Tool has been built using the transformation-based graphical tool building
framework METAclipse \cite{bib_metaclipse}. The functionality of the graphical tool is specified using model transformations in METAclipse. 
MOLA Tool has been built using MOLA itself. As a consequence a MOLA transformation definition is stored as a model 
and we can operate with it as with an ordinary model. Thus, we can use MOLA to generate MOLA. 
The abstract syntax of MOLA language has been used. 
See the MOLA metamodel in the reference
manual \cite{refman}.

Figures \ref{fig:hot1} and \ref{fig:hot2} show the higher-order transformation
written in MOLA generating the procedure \newline
\verb|FindOwnerOfTypeClassMethod| which implements the more complex task described in the previous section. See \url{http://mola.mii.lu.lv/img/generated.png} 
for an image of the procedure. Two additional parameters are added to help
finding the information required for implementing extensions of the case study.  
A normal switch case and a catch block containing the element are found.

Although it is possible to generate MOLA using MOLA itself, we think a better
option would be to use Template MOLA \cite{tmola} - a graphical
template language for transformation synthesis. Template MOLA uses the concrete
syntax of MOLA to specify MOLA elements being generated in a more readable way.
However, the tool for Template MOLA has not been  built properly yet.

Though generated MOLA procedure is in the abstract syntax, MOLA Tool generates the concrete syntax automatically 
and uses GraphViz\footnote{\url{http://www.graphviz.org/}} \emph{dot} for auto layout. 
Thus, the generated procedures can be viewed and edited further using MOLA Tool
as ordinary MOLA procedures. 
Development of higher-order transformation does not differ from development of ordinary MOLA transformation - the same MOLA Tool is used. However the technical space where transformations are
executed is different from EMF. Although the METAclipse framework is based
on Eclipse technologies, models are stored and transformations run on
the metamodel-based repository MII\_REP. Therefore the higher-order
transformation is executed directly on the repository used by MOLA Tool.
                                                                                                                                                                                      
\section{MOLA Transformation for the Reengineering Challenge}
\label{sec:sol}  

The MOLA solution has been shown in Figure \ref{fig:solution} (see Appendix).
The first rule finds the class named State. If the class exists, then the state machine is created. The creation is denoted using red color for class elements and 
association links in a rule. If the State class does not exist, then the
transformation terminates. 
Next the foreach loop is used to iterate through all non-abstract classes (see
A in Figure \ref{fig:solution}). Call to the \verb|isClassSubclassOf| procedure
is used to determine whether a class is a subclass of the \verb|State| class. If the class is a subclass then a state instance is created and put into the state machine. Additionally a traceability link is created.

The next foreach loop is used to create transitions. At first the core task
has been solved (see B in Figure \ref{fig:solution}). Every method call
conforming to the task description (\verb|State.Instance().activate()|) is handled by the loop. Call to the \verb|FindOwnerOfTypeClassMethod| method returns the owning class method (\verb|@own|), owning normal switch case (\verb|@ownNSC|) and owning
catch block (\verb|@ownCB|). When the owning method has been obtained,
the corresponding class and state can be found. In the same rule a transition is
created. 

The solution of the first extension starts with the next rule (see C in the
Figure \ref{fig:solution}). According to the description of the task the trigger
name for transitions whose activation occurs outside the \verb|run| method is equal to the methods name. If this
condition is satisfied, then the trigger name is set. Otherwise the second case should be examined. If the pattern of a MOLA rule fails, then the next statement reached by flow labeled \verb|{ELSE}| is executed. In our case, it is a rule checking, 
whether an owning normal switch case exists. If it exists, then the enumeration constant is found and set as a trigger name. 
Otherwise the third case should be examined. If an owning catch block exist, then the corresponding exception class 
is located and the trigger name set to the name of the class. Otherwise the trigger name is set to ``- -''. 

The two last rules of the loop solve the second extension (see D in Figure
\ref{fig:solution}). A call to the send method is searched in the statement list container owning the activation call. If such a method call is found then the
action names of the transitions are set to the name of enumeration constant
passed as an argument to the call. Otherwise the name of the action is set to
``- -''.    
   
\section{Conclusions}
\label{sec:dis}

In this paper the MOLA solution to the Reengineering Challenge has been
described. MOLA solution implements the core task as well as both extensions.
Since MOLA language lacks means how to deal with recursive, generic patterns in a concise and elegant way, 
an approach involving model transformation generation (higher-order transformations) has been proposed (HOT approach). 
The HOT approach allows to deal with complex 
situations when a model transformation language lacks these constructs. 

The main advantages of the HOT approach are:
\begin{itemize}
\item There is no need to change the implementation of a language to introduce
the desired functionality. We have added a procedure which finds a class
method owning the given arbitrary Java element.
\item HOTs are reusable - the same transformation can be used in another model transformation project.
\item HOTs are flexible - if some changes are needed to the functionality of the generated model 
transformation it can be easily added by changing the HOT definition or even adding the new functionality 
manually to the generated code. In our solution the generated procedure searches
for an owner (normal switch case or catch block) additionally to the class
method.
\end{itemize} 
 
The main disadvantage of the HOT approach is that it requires a deep knowledge of the language being generated. 
MOLA metamodel (abstract syntax) should be familiar to a developer of a HOT.
This leads to the overspecification. It is the main
reason why the solution was awarded with a lower score for the conciseness. We
believe that the graphical versus textual languages issue has also played a
significant, but not the main, role (graphical languages got the lowest scores for the conciseness). A HOT specification using the concrete syntax of the
transformation language being generated would be a great improvement. Another issue is the maturity of HOT tools. In our case running higher-order transformation requires a deep understanding of the architecture of MOLA Tool.

The MOLA solution has been tested on the models provided by the case author. The
execution of transformation on the simple model took in total 1280 ms on a
virtual machine within the {SHARE}
environment\cite{molashare}. It includes 280 ms loading model, 170 ms copying to
the intermediate representation (having associations navigable both ways), 720 ms actually
running the transformation, less than 1 ms extracting the target model from
intermediate representation and finally 110 ms saving the target model. The
execution of the transformation on the medium model took 1600 ms in total. The execution of transformation on 
the big model failed due to memory consumption problems of MOLA runtime
environment. The problem is being fixed by improving the algorithm and
fixing bugs in the runtime environment.

\subsubsection*{Acknowledgments.} This work has been partially supported by the European Social Fund within the project ``Support for Doctoral Studies at University of Latvia''.

\bibliographystyle{eptcs}
\bibliography{ttc2011}

%
%
%
%
%


\newpage
\section{Appendix}

Figure \ref{fig:isclasssubclassof} (on page 2) provides an example of a model
transformation procedure written in MOLA language. The \verb|isClassSubclassOf| procedure answers whether a class is a subclass of another class. It is a simple case 
for non-local matching of model elements. There are three parameters for the procedure 
(white boxes in the upper part of the procedure). The first two parameters are classes to be examined. The third parameter 
is an in-out parameter where the result (true or false) is stored.

The procedure contains a foreach loop iterating over all superclasses of the given class. If a superclass is the given one, then the result is \emph{true}. 
Otherwise the procedure is called recursively passing the superclass as the first parameter. 
If no class in the inheritance hierarchy corresponds to the given class, 
then the result is \emph{false}.

\begin{figure}[h] 
\centering
\includegraphics[width=\columnwidth]{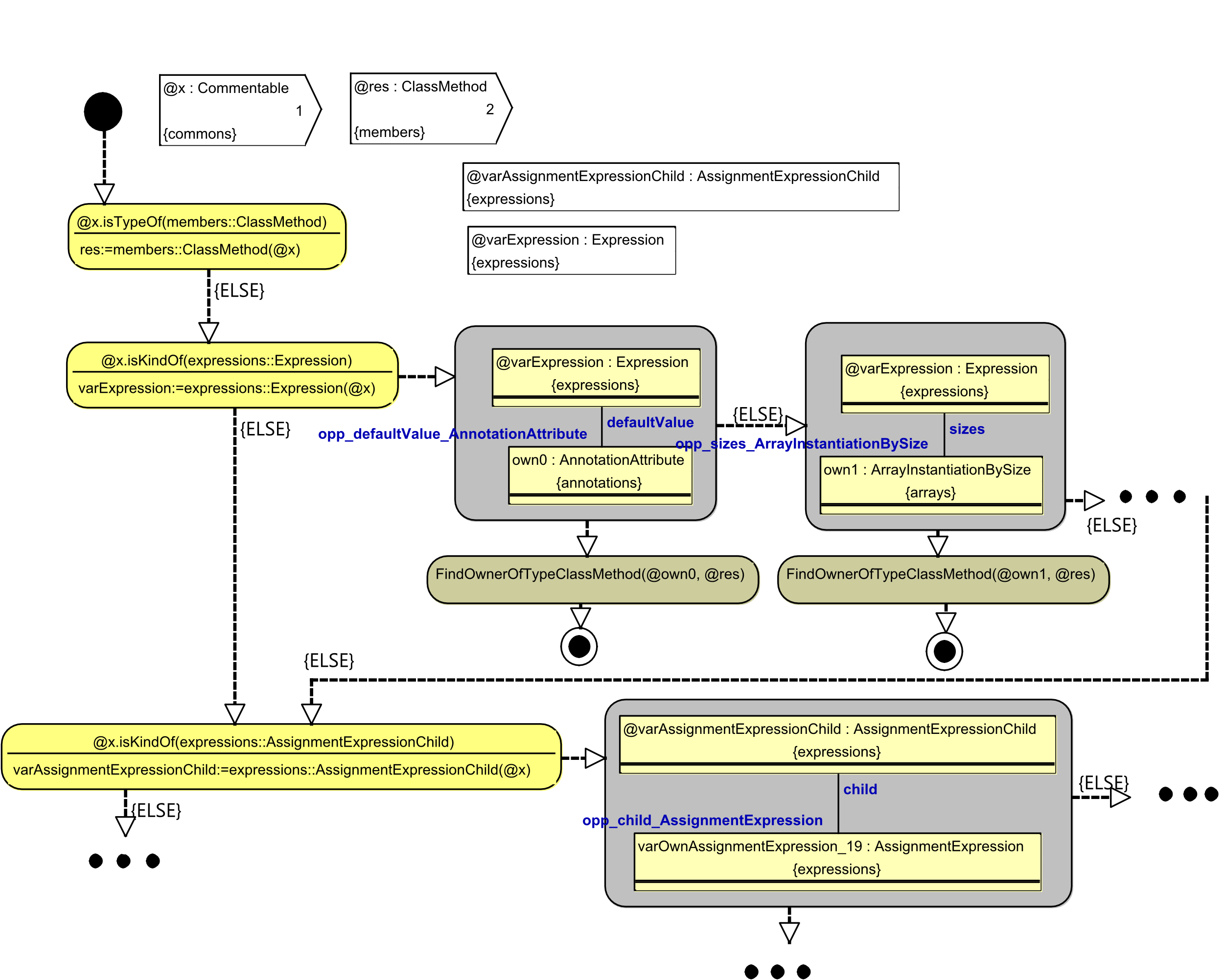}
\caption{Excerpt of the FindOwnerOfTypeClassMethod procedure}
\label{fig:findowner}
\end{figure}

Figure \ref{fig:findowner} shows an excerpt of a procedure finding an owning class method for an arbitrary Java element. 
The procedure has two parameters (white boxes in the upper part of the Figure
\ref{fig:findowner}) - the first one is the Java element corresponding to
\newline \verb|commons::Commentable| class (every element in
the metamodel extends this class), the second is an in-out parameter containing
the result - the owning class method. If the passed argument  is already a class
method, it is the answer and execution of the procedure ends. Otherwise the type
of the argument should be determined (actually the kind of the argument, not
the exact type). It is done using a chain of \emph{text statements} (yellow
rounded rectangles) linked by \verb|{ELSE}| flows. We are interested only in
those types which have containment associations. When the type of the argument
is known, we can check whether the argument has a containment link. If so, then
the procedure is called recursively passing the container as an argument.
Otherwise we try to determine whether the argument corresponds to another type. The recursive calls stop, when the owning class method is found or the root of the containment hierarchy has been reached.

\begin{figure}
\centering
\includegraphics[width=\columnwidth]{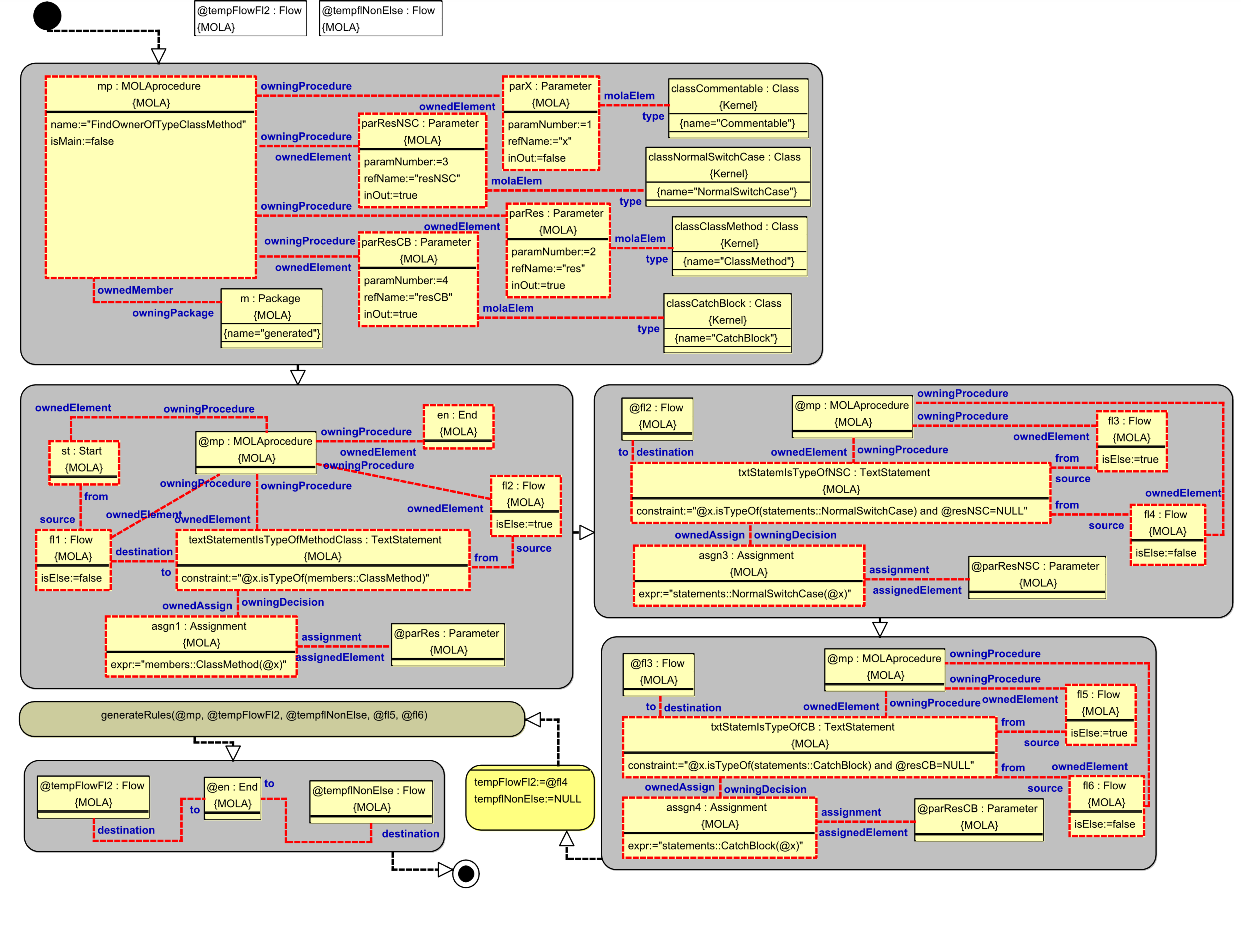}
\caption{Higher-Order Transformation generating the
fixed part of FindOwnerOfTypeClassMethod procedure}
\label{fig:hot1}
\end{figure}

Figure \ref{fig:hot1} shows the first part of higher-order transformation. Two
upper MOLA rules generate the header and the text statement of the procedure checking if the argument is a class method. 
The next two rules generate two text statements checking if the argument is a normal switch case or a catch block. 

\begin{figure}
\centering
\includegraphics[width=\columnwidth]{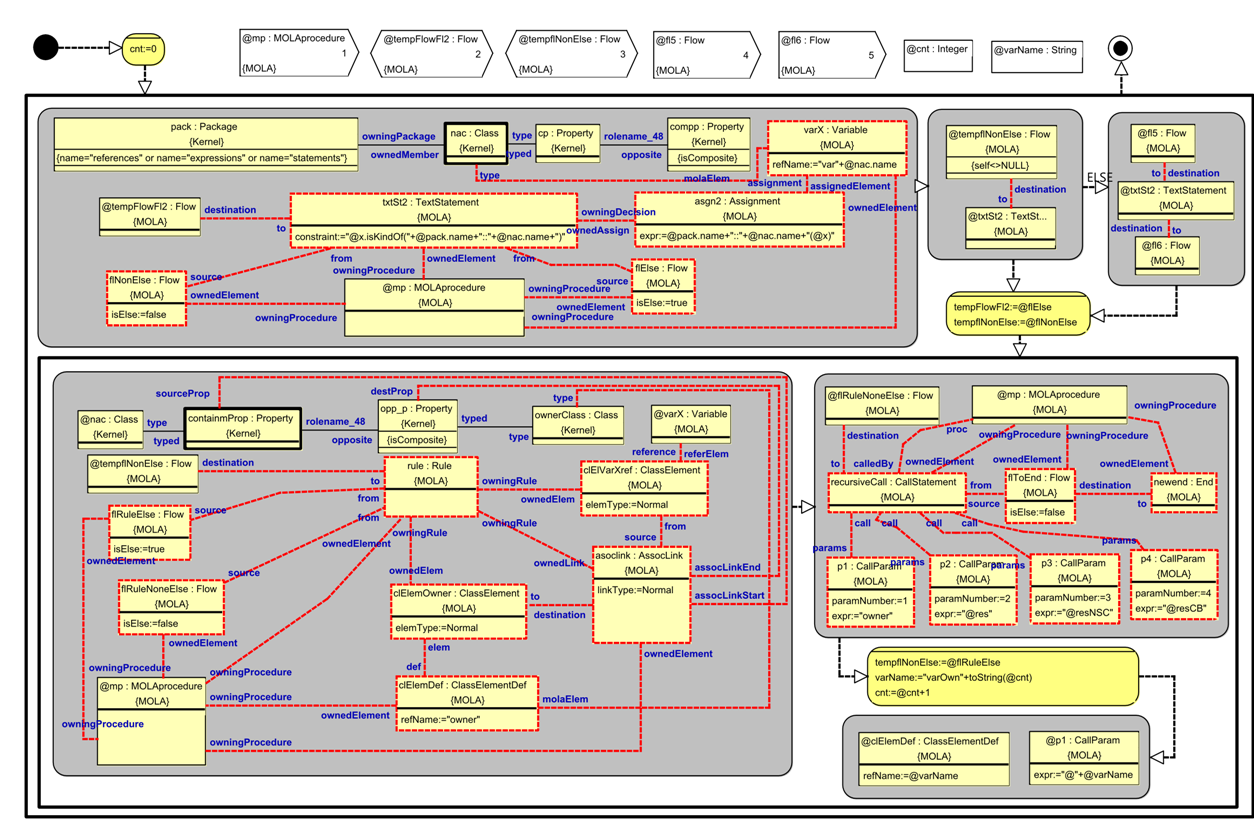}
\caption{Higher-Order Transformation generating the
variable part of FindOwnerOfTypeClassMethod procedure}
\label{fig:hot2}
\end{figure}

The next statement is a foreach loop (see Figure \ref{fig:hot2}). The loop
iterates over all metamodel classes from the \verb|references|, \verb|expressions| and \verb|statements| packages having an outgoing composition. 
For each class a text statement checking if the argument is an instance of this class is generated. 
The nested foreach loop goes through all composite associations of the class and generates a 
MOLA rule and a call statement for each. All other MOLA rules and text statements have been used 
to create appropriate flows between generated MOLA statements.

\begin{figure}
\centering
\includegraphics[width=\columnwidth]{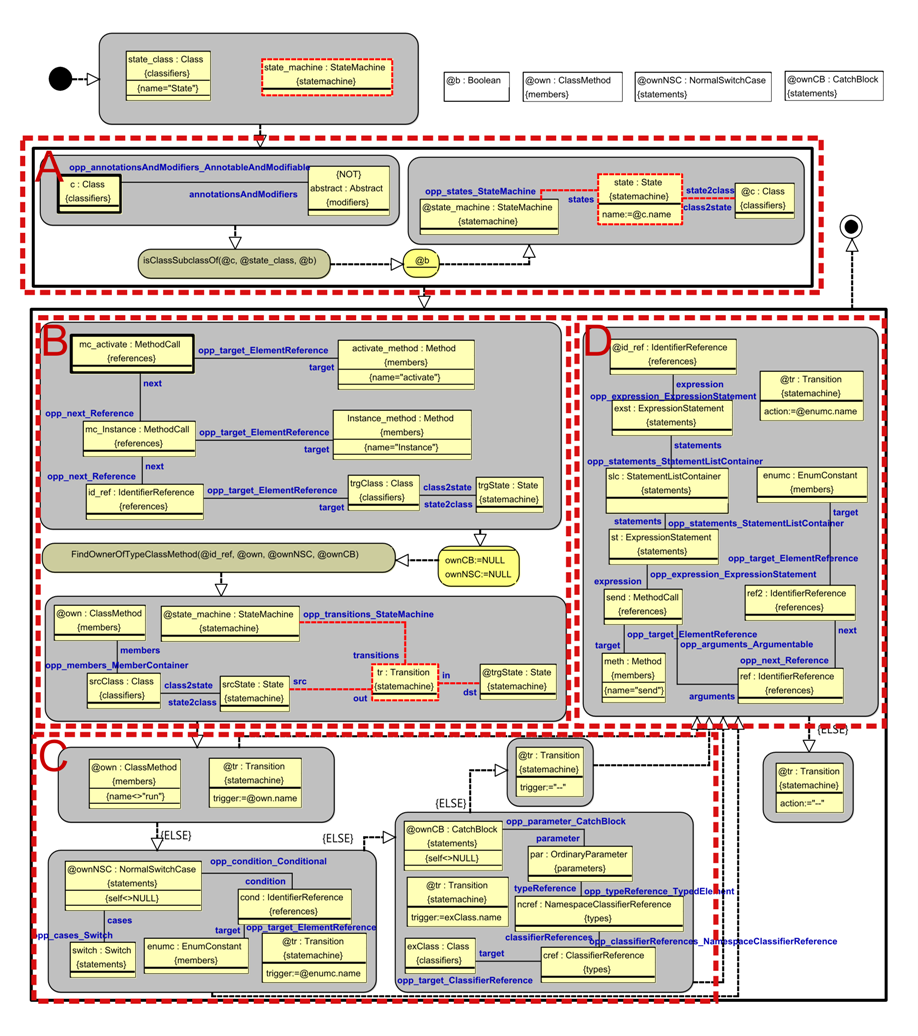}
\caption{The main part of the MOLA solution}
\label{fig:solution}
\end{figure}
\end{document}